# Nonlinear ultrafast modulation of the optical absorption of few-cycle terahertz pulses in n-doped semiconductors

L Razzari[(1,2)]*, F H Su[(3)], G Sharma[(1)], F Blanchard[(1)], A Ayesheshim[(3)], H-C Bandulet[(1)], R Morandotti[(1)], J-C Kieffer[(1)], T Ozaki[(1)], M Reid[(4)], and F A Hegmann[(3)†]

[(1)] *INRS-EMT, Advanced Laser Light Source, Université du Québec, Varennes, Québec J3X 1S2, Canada*

[(2)] *Dipartimento di Elettronica, Università di Pavia, via Ferrata 1, 27100 Pavia, Italy*

[(3)] *Department of Physics, University of Alberta, Edmonton, Alberta T6G 2G7, Canada*

[(4)] *Department of Physics, University of Northern British Columbia, Prince George, British Colombia V2N 4Z9, Canada*

* corresponding author; luca.razzari@emt.inrs.ca

[†] hegmann@phys.ualberta.ca

**Abstract:**

We use an open-aperture Z-scan technique to show how intense few-cycle terahertz pulses can experience a nonlinear bleaching of absorption in an n-doped semiconductor due to terahertz-electric-field-driven intervalley scattering of electrons in the conduction band. Coherent detection of the transmitted terahertz pulse waveform also allows the nonlinear conductivity dynamics to be followed with sub-picosecond time resolution. Both the Z-scan and time-domain results are found to be in agreement with our theoretical analysis.

Ultrafast nonlinear processes have been extensively studied in the visible and near-infrared [1], but have remained basically unexplored in the terahertz spectral region because of the lack of sources delivering high-energy, few-cycle terahertz pulses. Nowadays, these sources are becoming available [2-6], allowing the possibility of investigating novel and exciting aspects of radiation-matter interaction.

Over the past 25 years, few studies have explored the nonlinear dynamics of semiconductors using terahertz (THz) sources with microsecond or nanosecond pulse durations [7-9]. Only very recently, however, coherently-detected intense terahertz pulses with picosecond pulse durations have become available to study the nonlinear response of different physical systems [10-13]. For example, Hebling et al. [10] have explored ionic nonlinearities in $LiNbO_3$ crystals, observing significant lattice anharmonicities induced by strong terahertz electric fields. Gaal et al. [11] have reported a nonlinear electronic response in n-type GaAs, where a long-lived coherent THz emission, centered around 2 THz, and driven by intense THz pulses has been observed and attributed to stimulated emission from impurities exhibiting a population inversion. These two studies are good examples of how broad can be the interest associated to ultrafast nonlinear spectroscopy in the terahertz region, since it can give access to either ionic or electronic information, depending on the properties of the excited system. In principle, this new spectroscopy can also explore drift-velocity-based nonlinearities of free carriers in semiconductors, since the low terahertz photon energy makes multiphoton interband effects negligible even in narrow-bandgap semiconductors [14]. This kind of free-carrier nonlinearities have been studied in the past in the microwave region [15,16], but the availability of ultrashort intense terahertz pulses can now unveil the dramatic dynamics of these processes with an unprecedented time resolution.

In order to explore these new nonlinearities we have decided to borrow one of the most common and straightforward nonlinear optical characterization techniques, namely open-aperture Z-scan, and apply it to the THz regime [17]. It consists in scanning the sample transmission through the focus of an intense optical beam (see Figure 1a) and thus it gives access to nonlinear (i.e. intensity dependent) transmission properties. Widely used in multiphoton absorption studies [18,19], the Z-scan technique has proven to be effective even for characterizing saturable absorbers [20]. A large-aperture ZnTe optical rectification source that we have recently developed in our labs [2] is used as the source of high power THz radiation. It delivers picosecond terahertz pulses in the frequency range of ~ 0.1 to 3 THz with µJ-level energies, at a repetition rate of 100 Hz. Figure 1(b) shows an example of the temporal profile of the THz pulses produced by our source, and the inset shows the corresponding power spectrum. These pulses are focused, by means of an off-axis parabolic mirror, down to a spot diameter of 1.6 mm ($1/e^2$ value of the intensity profile, see Fig. 1(c)) at the focal position of the Z-scan ($z = 0$). The THz beam profile is found to be well fitted by a Gaussian shape.

With regard to the sample, we have chosen as a test-bed a thin film of a heavily doped direct band-gap semiconductor widely used in optoelectronics, namely Indium Gallium Arsenide. More specifically, it consists of a 500 nm-thick n-type $In_{0.53}Ga_{0.47}As$ epilayer with a doping concentration of approximately 2 x $10^{18}$ $cm^{-3}$ grown by metal-oxide chemical vapor deposition on a lattice-matched, 0.5-mm-thick semi-insulating InP substrate. At low excitation levels, the sample transmits approximately 3% of the incident energy; this strong drop in transmission has been found to be mainly due to the high conductivity of the epilayer, since measurements of the InP substrate alone have shown an overall transmission (including absorption and reflection losses) of about 60%.

Figure 2(a) shows the results of a Z-scan experiment on the sample using a standard pyroelectric detector (Coherent Molectron J4-05) to detect the total transmitted THz pulse energy. As one can see in Fig. 2(a), by illuminating the sample with 0.8 µJ terahertz pulses (peak electric field of approximately 200 kV/cm), a significant enhancement in the transmission is observed near the focus of the Z-scan relative to the transmission away from the focus. Fig. 2(a) shows also that this dramatic bleaching effect *is not observed* when the same scan is carried out on a bare InP substrate. In the only other terahertz Z-scan experiment performed to date (using an undoped InSb sample [13]), the authors observed an absorption *enhancement* due to impact ionization effects rather than a *bleaching* due to intervalley scattering . A nonlinear bleaching of terahertz pulse absorption, similar to the one reported here, was reported in n-doped Ge and GaAs by a pioneering work [8], and was attributed to a THz-electric-field-induced scattering of carriers into satellite valleys of the conduction band. In these satellite valleys, electrons acquire a significantly higher effective mass, reducing the macroscopic conductivity of the sample and thus increasing the transmission. However, the 40 ns pulse duration of the THz pulses [8] did not allow the ultrafast dynamics of the nonlinear mechanism to be resolved. The very recent new-generation few-cycle terahertz sources, and coherent detection techniques based on electro-optical sampling, can now shed light on the dynamics and the real nature of this phenomenon.

Figure 2(b) shows the temporal profiles of transmitted terahertz pulses at different z-positions. Taking the time integral of the modulus squared of these traces, one indirectly retrieves a quantity proportional to the transmitted energy (Fig. 2(c)), which is consistent with the direct energy measurement shown in Fig. 2(a). Figure 2(b) shows no significant temporal shift between the transmitted pulses taken at different positions relative to the focus, so that changes in the imaginary part of the conductivity are negligible in this set of measurements. One can also see that the transmission enhancement is not uniform in time, suggesting that the underlying nonlinear process has dynamical features on a time scale comparable to (or slightly larger than) the THz pulse duration.

The dynamics of the bleaching process are evident when the difference in electric field transmission between the actual position in the Z-scan and a position far away from the focus for each peak of the terahertz pulse is plotted as in Fig. 2(d). The curves show an initial increase in transmission over a 1 ps time interval (with a peak at t=2.2 ps in Fig. 2(d)) followed by a slower decay over approximately a 2 ps time scale. These features can be explained as follows. Free carriers in the $\Gamma$ valley are accelerated by the terahertz electric field during each of its oscillations. When the carriers acquire enough kinetic energy to overcome the nearest intervalley separation, they may scatter into an upper valley (i.e. the L-valley) where the effective mass is higher. The carrier mobility is therefore lower in the satellite valleys, which reduces the overall conductivity of the film. Since the transmission of the THz pulse depends on the conductivity of the InGaAs film, the transmission is enhanced when carriers are scattered to the upper satellite valleys. The electrons in this upper valley will then have a finite probability of scattering back to the $\Gamma$ valley where the effective mass is smaller, thereby increasing again the conductivity of the film back to its original value. This in turn results in a drop in the THz transmission as carriers scatter back to the Γ valley with a time constant given by the L → Γ intervalley scattering time. The closest upper valley in In$_{0.53}$Ga$_{0.47}$As is the L valley ($\Delta_{\Gamma L}$=0.55 eV). The effective masses in the two valleys are m*$_{\Gamma 0}$=0.03745m$_e$ and m*$_{L0}$=0.26m$_e$ respectively, and the L-Γ intervalley relaxation time has been previously estimated to be about 3.1ps [21], which is consistent with the decay dynamics observed in Fig. 2(d). It is important to note that a frequency-domain analysis of the process yields misleading results, since the chirp in the THz pulse waveform, which is always present in ultrashort terahertz pulses, creates artificial frequency dependencies of the bleaching process since different parts of the waveform (and therefore different frequency components) have different amplitudes. Thus, it is necessary to study nonlinear terahertz pulse interactions in the time domain.

Electric-field-driven intervalley scattering is a well-known mechanism in high-field transport physics, and it is known to cause negative differential resistance and Gunn oscillations in direct bandgap semiconductors [22]. The critical DC field required to excite these phenomena in In$_{0.53}$Ga$_{0.47}$As is usually in the range 2.5-4 kV/cm [23]. Because of the high frequency of the exciting electric field used here, the critical field associated with our experiments can be higher than its DC counterpart, since during one terahertz pulse oscillation the electrons can acquire energies higher than the intervalley separation (a phenomenon known as velocity overshoot [22]), before scattering into the L valley. This is consistent with our experimental findings, where the effect vanishes rapidly as the peak electric field inside the epilayer drops below 14 kV/cm (position z=±4 mm in the Z-scan).

The terahertz transmission of our sample, which can be idealized as a thin conducting sheet with thickness $d$ on an insulating substrate with index $n$, can be expressed as [24]:

$$E_t = \frac{1}{Y_0 + Y_s}(2Y_0 E_i - Jd), \qquad (1)$$

where $E_t$ and $E_i$ are the transmitted and incident fields, respectively, $Y_0 = (377\ \Omega)^{-1}$ and $Y_s = nY_0$ are the free-space and substrate admittances, and $J$ is the current density in the film. During the absorption bleaching process, the transmitted field $E_t$ accelerates the electrons in the conducting InGaAs layer and drives the population transfer between the different valleys of the conduction band, thus affecting the current density $J$ in Eq. (1), which in turn modifies the transmitted field. This feedback is responsible for the rich and somehow surprising dynamical features associated with the effect under investigation. In particular, a relatively simple mathematical model based on the Drude model is proven to be adequate enough to describe the collective motion of the electrons within the film in response to the THz electric field. The change in electron populations in the $\Gamma$ and L valleys is determined by the intervalley scattering rates. The L-$\Gamma$ transfer rate $\tau^{-1}_{L\Gamma}$ is kept constant [25], while the $\Gamma$-L scattering rate $\tau^{-1}_{\Gamma L}$ is a function of the average kinetic energy of the electrons in the $\Gamma$ valley [26]. Such transfer rate is zero at low energies but starts to increase rapidly at a threshold value $\varepsilon_{th}$ to a maximum value $\tau^{-1}_{\Gamma L0}$ at high energies. Band nonparabolicities were also included in the simulation [27]. Figure 3 shows how well this relatively simple model can describe *both* the z-scan *and* the time-domain experimental results, strongly suggesting that the intervalley scattering mechanism is responsible for the observed nonlinear process. Table I summarizes the parameters used in the simulation [27-30]. As one can see, the threshold energy value that has to be assumed to fit the experimental data (0.13eV) is lower than the intervalley separation commonly found in literature (0.55 eV). This is most probably due to the fact that our simplistic model does not take into account the electron temperature, which is usually a significant term in the electron energy equation. However, the ability of the model to properly describe the dynamics of the process is demonstrated by the maximum drift velocity predicted at the focal position, which is around $9\times10^7$ cm/s. Calculations made by other groups have shown that this value, which is limited by intervalley scattering in InGaAs, can reach $10^8$ cm/s at a time of 200 fs after application of a 20 kV/cm electric field [30]. Since the effective mass of the electron in InGaAs is about 0.04, then this maximum velocity overshoot value corresponds to a kinetic energy of about 0.11 eV, which is much less than the $\Gamma$-L energy separation of 0.55 eV but very close to the threshold energy of 0.13 eV obtained from our fits to the data. The fitting procedure allows one to quantify the scattering rates of heavily-doped InGaAs, as studied here. The $\Gamma$-L intervalley scattering rate is found to be approximately $3.33 \times 10^{13}$ s$^{-1}$, close to the value measured in GaAs [29], while the L–$\Gamma$ intervalley relaxation rate is found to be approximately $2.50\times10^{11}$ s$^{-1}$, similar to that found in [21].

Using this model we can also describe the rapid transmission change along the Z-scan (Fig. 3(c)). The process is highly nonlinear, as seen in Fig. 3(d), where the normalized transmission is plotted with the estimated incident field strengths along the Z-scan. It should be noted that this nonlinear behavior cannot be reproduced by any two-level-system saturation model, like the one reported in Ref. 20, and it offers further evidence of the intervalley bleaching mechanism discussed here. Our THz Z-scan measurements further illustrate how dynamic processes can be extracted from the changes in transmission of a *single* THz pulse.

In conclusion, the newly opened field of THz ultrafast nonlinear spectroscopy has proven to be a surprisingly effective and simple method for the characterization of drift-velocity-based nonlinearities in semiconductors. Significant transmission changes due to intervalley scattering processes induced by the electric field of the THz pulse have been observed over picosecond time scales.

Beside the fundamental relevance associated to one of the first demonstrations of ultrafast nonlinear optics in the THz regime, the present work could find several exciting applications, including the realization of a semiconductor-based terahertz saturable absorber.

We wish to acknowledge financial support from NSERC, NSERC Strategic Projects, and INRS. L.R. would also like to acknowledge a Marie Curie Outgoing International Fellowship (contract n. 040514).

**Figure 1**

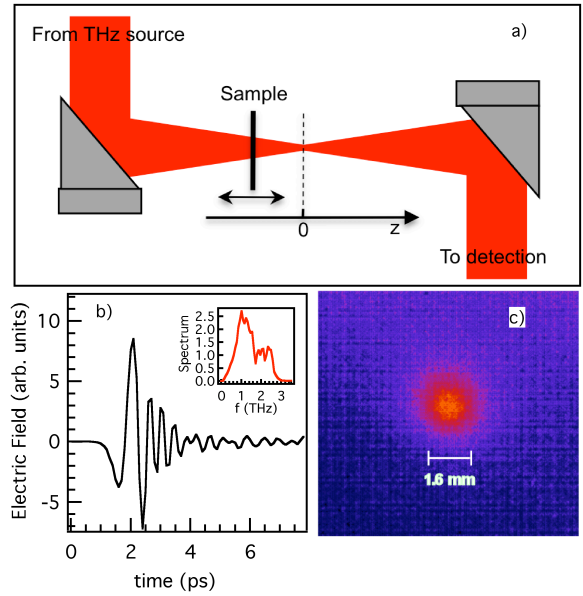

**Fig. 1.** Experimental setup. **(**a) Schematic of the experimental setup. (b) Electric field profile of the terahertz beam emitted by the ZnTe optical rectification source. Inset: power spectrum of the THz pulse. (c) Pyroelectric image of the terahertz spot-size at the focus (z=0), with a $1/e^2$ diameter of 1.6 mm.

**Figure 2**

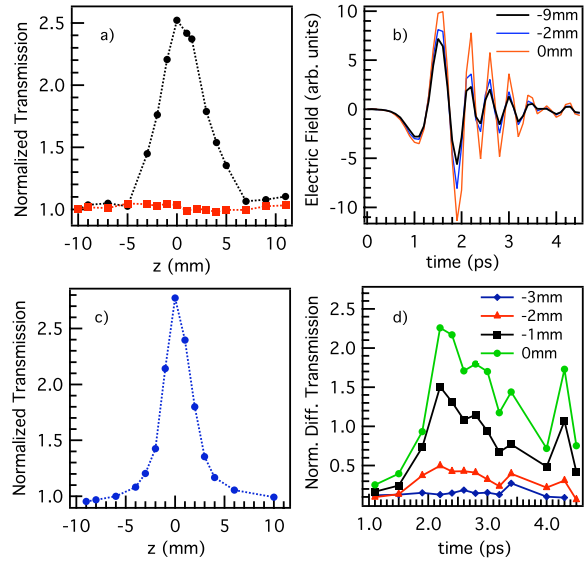

**Fig. 2.** Experimental results. (a) Z-scan normalized transmission of total THz pulse energy measured with a pyroelectric detector after the sample; black curve: InGaAs epilayer on an InP substrate. Red curve: InP substrate alone. (b) Transmitted THz pulse electric field for different positions of the Z-scan. (c) Normalized transmission of the time integral of the modulus squared of the transmitted electric field as a function of the z position along the scan. (d) Normalized electric field differential transmission as a function of time for different z position along the scan. Note that the initial positive slope is related to the THz pulse duration and the population rate, while the negative slope is indicative of the carriers decay time.

**Figure 3**

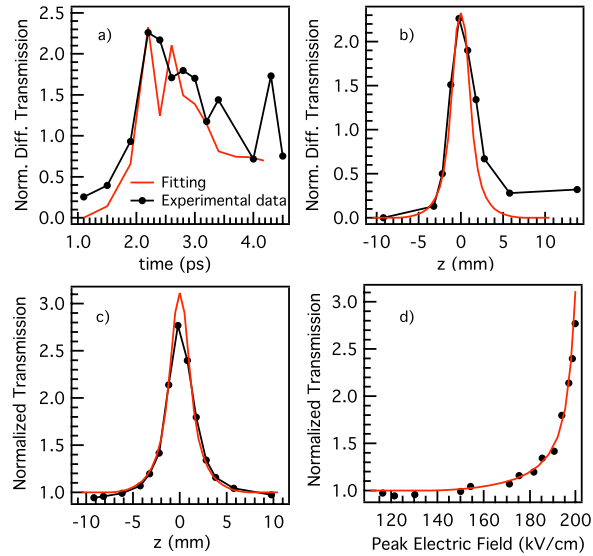

**Fig. 3.** Comparison to model. (a) Normalized electric field differential transmission as a function of time at the focus of the Z-scan. (Red line: model; Black line: experiment). (b), Peak value (t=2.2ps) of the normalized electric field differential transmission as a function of the z position along the scan. (c) Comparison to data from Fig. 2(c). (d) Incident electric field dependence of the normalized energy transmission.

**Table I**

| Parameters | symbols | simulation parameters | References |
|---|---|---|---|
| Effective mass ratio | $m^*_{\Gamma 0}$ | | Γ valley: 0.03745 [27] |
| | $m^*_{L0}$ | | L valley: 0.26 [27] |
| Nonparabolicity factor (eV$^{-1}$) | $a_\Gamma$ | | Γ valley: 1.33 [27] |
| | $a_L$ | | L valley: 0.59 [27] |
| Refractive index of InP | n | | 3.1 [28] |
| Threshold energy (eV) | $\varepsilon_{th}$ | 0.13 | 0.25 [30] (a) |
| Smooth parameter | b | 0.57 | |
| Intravalley scattering rate (s$^{-1}$) | $\tau_\Gamma$ | Γ valley: 1.0 x10$^{13}$ | |
| | $\tau_L$ | L valley: 1.67 x10$^{13}$ | |
| Γ-L intervalley scattering rate (s$^{-1}$) | $\tau^{-1}_{\Gamma L0}$ | 3.33 x10$^{13}$ | 2.50x10$^{13}$ [29] (b) |
| L-Γ intervalley scattering rate (s$^{-1}$) | $\tau^{-1}_{L\Gamma}$ | 2.50 x10$^{11}$ | 3.23x10$^{11}$ [21] |

a) We estimated that the threshold energy is lower than 0.25eV according to the Monte Carlo calculation reported on doped In$_{0.53}$Ga$_{0.47}$As in reference [30].

b) The Γ−L intervalley scattering rate of GaAs reported in reference [29] is about 2.50x10$^{13}$ s$^{-1}$.

**Table 1.** Model parameters.